# Very Fast Watermarking by Reversible Contrast Mapping

Dinu Coltuc and Jean-Marc Chassery

*Abstract*—Reversible contrast mapping (RCM) is a simple integer transform that applies to pairs of pixels. For some pairs of pixels, RCM is invertible, even if the least significant bits (LSBs) of the transformed pixels are lost. The data space occupied by the LSBs is suitable for data hiding. The embedded information bit-rates of the proposed spatial domain reversible watermarking scheme are close to the highest bit-rates reported so far. The scheme does not need additional data compression, and, in terms of mathematical complexity, it appears to be the lowest complexity one proposed up to now. A very fast lookup table implementation is proposed. Robustness against cropping can be ensured as well.

*Index Terms*—Difference expansion, embedding bit-rate, reversible contrast mapping (RCM), reversible watermarking.

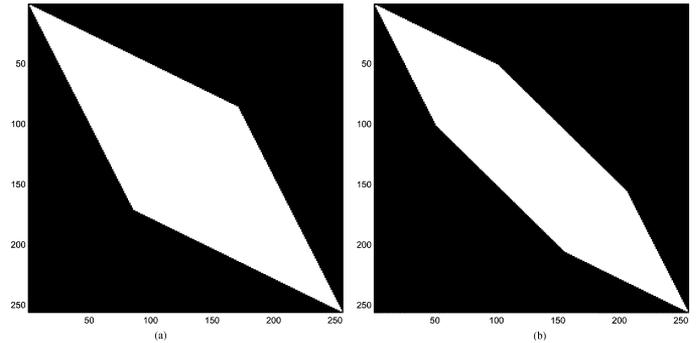

Fig. 1. Transform domain $D$: (a) without and (b) with control distortion.

## I. INTRODUCTION

MOST of the reversible watermarking approaches proposed so far incorporate a lossless data compression stage [1]–[4]. The use of an elaborate data compression stage increases the mathematical complexity of the watermarking. There are some watermarking schemes that do not rely on additional data compression, as for instance, the circular histogram interpretation schemes [5], but they have the drawback of a low embedding capacity.

In this letter, we discuss a spatial domain reversible watermarking scheme that achieves high-capacity data embedding without any additional data compression stage. The scheme is based on the reversible contrast mapping (RCM), a simple integer transform defined on pairs of pixels. RCM is perfectly invertible, even if the least significant bits (LSBs) of the transformed pixels are lost. The data space occupied by the LSBs is suitable for data hiding.

The basic RCM watermarking scheme was introduced in [6]. Here, a modified version that allows robustness against cropping is proposed. The control of distortions introduced by the watermarking is investigated as well. The mathematical complexity of the RCM watermarking is further analyzed, and a very low cost implementation is proposed. Finally, the RCM scheme is compared with Tian's difference expansion scheme [3] with respect to the bit-rate hiding capacity and to the mathematical complexity. It is shown that the RCM scheme provides almost similar embedding bit-rates when compared to the difference expansion approach, but it has a considerably lower mathematical complexity.

## II. REVERSIBLE CONTRAST MAPPING

Let $[0, L]$ be image graylevel range ($L = 255$ for eight-bit graylevel images), and let $(x, y)$ be a pair of pixels. The forward RCM transforms pairs of pixels into pairs of pixels

$$x' = 2x - y, \quad y' = 2y - x. \tag{1}$$

To prevent overflow and underflow, the transform is restricted to a subdomain $D \subset [0, L] \times [0, L]$ defined by the equations

$$0 \leq 2x - y \leq L, \quad 0 \leq 2y - x \leq L. \tag{2}$$

As shown in Fig. 1(a), $D$ is a rhombic domain located along the diagonal of $[0, L] \times [0, L]$.

The inverse transform is defined as follows:

$$x = \left\lceil \frac{2}{3}x' + \frac{1}{3}y' \right\rceil, \quad y = \left\lceil \frac{1}{3}x' + \frac{2}{3}y' \right\rceil \tag{3}$$

where $\lceil a \rceil$ is the ceil function (the smallest integer greater than or equal to $a$).

As stated in Section I, the pair forward-inverse transform should give exact results, even if the LSBs of the transformed pixels are lost. If $x'$ and $y'$ are not changed, (3) exactly inverts (1), even without the ceil functions. By watermarking, the LSBs of $x', y'$ are lost. Let us set to "0" the LSBs of $x'$ and $y'$. It immediately appears that if the LSB of $x'$ was "1," the values inside the ceil functions for the computation of $x$ and $y$ decrease with 2/3 and 1/3, respectively. Similarly, if the LSB of $y'$ was "1," the corresponding values decrease with 1/3 (for the computation of $x$) and 2/3 (for the computation of $y$). Except when both LSBs are "1," the ceil function recovers the correct results. An LSB of "1" means an odd integer number. From (1), it follows that

Manuscript received April 13, 2006; revised August 3, 2006. This work was supported by ANR, France, under project ARA-TSAR. The associate editor coordinating the review of this manuscript and approving it for publication was Dr. Mauro Barni.

D. Coltuc is with the Valahia University of Targoviste, Bucharest, Romania, and also with LIS—National Polytechnic Institute of Grenoble, Grenoble, France, (e-mail: coltuc@valahia.ro.).

J.-M. Chassery is with the LIS—National Polytechnic Institute of Grenoble, Grenoble, France (e-mail: jean-marc.chassery@lis.inpg.fr).

Digital Object Identifier 10.1109/LSP.2006.884895

$(x', y')$ are both odd numbers only if $(x, y)$ are odd numbers, too. To conclude, on $D$ without the set of odd pairs, the inverse RCM transform performs exactly, even if the LSBs of the transformed pairs of pixels are lost.

The forward transform should not introduce visual artifacts. By taking the sum and the difference of (1), one gets $x' + y' = x + y$ and $x' - y' = 3(x - y)$, respectively. This means that RCM preserves the graylevel averages and increases the difference between the transformed pixels. Consequently, image contrast increases.

## III. REVERSIBLE WATERMARKING

The watermark substitutes the LSBs of the transformed pairs. At detection, in order to extract the watermark and to restore the original pixels, each transformed pair should be correctly identified. The LSB of the first pixel of each pair is used to indicate if a pair was transformed or not: "1" for transformed pairs and "0" otherwise.

The inverse RCM fails to recover the pairs $(x, y) \in D$ composed of odd values. Such pairs can be used as well for data embedding as long as they are correctly identified at detection. This can be easily solved by setting the LSB of the first pixel to "0." At detection, both LSBs are set to "1" and (2) are checked. If (2) are fulfilled, the pair was composed of odd pixels. In order to avoid decoding ambiguities, some odd pixel pairs should be eliminated, namely, those pairs located on the borders of $D$. The pairs subject to ambiguity are found by solving in odd numbers the equations: $2x - y = 1$, $2y - x = 1$, $2x - y = L$, and $2y - x = L$. For $L = 255$, there are only 170 such pairs. Let further $D_c$ be the domain of the transform without the ambiguous odd pixel pairs.

### A. Marking

The marking proceeds as follows.
1) Partition the entire image into pairs of pixels (for instance, on rows, on columns, or on any space filling curve).
2) For each pair $(x, y)$:
   a) If $(x, y) \in D_c$ and if it is not composed of odd pixel values, transform the pair using the (1), set the LSB of $x'$ to "1," and consider the LSB of $y'$ as available for data embedding.
   b) If $(x, y) \in D_c$ and if it is composed of odd pixel values, set the LSB of $x$ to "0," and consider the LSB of $y$ as available for data embedding.
   c) If $(x, y) \notin D_c$, set the LSB of $x$ to "0," and save the true value.
3) Mark the image by simple overwriting the bits identified in 2a and 2b with the bits of the watermark (payload and bits saved in 2c).

A different marking procedure is proposed in [6]. A map of transformed pairs and the sequence of LSBs for all nontransformed pairs are first collected. Then, the entire image LSB plane is overwritten by the payload and by the collected bit-sequences. The slightly modified procedure proposed in this letter provides robustness against cropping. The location map of the entire image is replaced by the LSB of the first pixel of each pair showing if the pair was transformed or not. Let us further consider that the saved LSB of a nontransformed pair is embedded into the available LSB of the closest transformed pair.

Thus, all the information needed to recover any original pixel pair is embedded into the pair itself or very close to it. In the case of cropping, except for the borders where some errors may appear, the original pixels of the cropped image are exactly recovered together with the embedded payload. For pixel pairing on row or column direction, there are no problems of synchronization. Some control codes should be inserted in the payload to validate watermark integrity.

### B. Detection and Original Recovery

Watermark extraction and exact recovery of the original image is performed as follows.
1) Partition the entire image into pairs of pixels.
2) For each pair $(x', y')$:
   a) If the LSB of $x'$ is "1," extract the LSB of $y'$ and store it into the detected watermark sequence, set the LSBs of $x'$, $y'$ to "0," and recover the original pair $(x, y)$ by inverse transform (3).
   b) If the LSB of $x'$ is "0" and the pair $(x', y')$ with the LSBs set to "1" belongs to $D_c$, extract the LSB of $y'$, store it into the detected watermark sequence, and restore the original pair as $(x', y')$ with the LSBs set to "1."
   c) If the LSB of $x'$ is "0" and the pair $(x', y')$ with the LSBs set to "1" does not belong to $D_c$, the original pair $(x, y)$ is recovered by replacing the LSB of $x'$ with the corresponding true value extracted from the watermark sequence.

### C. Data Hiding Capacity

Let $P$ be the total number of pairs, and let $T$ be the number of pairs with embedded information. The scheme provides $T$ bits of free space for data embedding. Besides the payload, the LSB of the first pixel of the other $P - T$ pairs should be stored, i.e., only $2T - P$ bits are available. The bit-rate provided by the scheme is

$$B = \frac{2T - P}{2P} \text{ bpp.} \qquad (4)$$

The scheme provides space for data embedding if at least half of the total number of pairs are transformed, i.e., $T > P/2$. The upper bound of the RCM scheme, $B_u$, is obtained when $T$, the number of pixel pairs with embedded data, is very high ($T \approx P$): $B_u = 0.5$ bpp.

In order to increase the data hiding capacity, multiple iterations of the algorithm are chained. For instance, after one iteration on rows, a second iteration on columns is done and so on. It should be noticed that, with each new iteration, the image distortion increases while the number of pairs belonging to $D_c$ decreases.

### D. Distortion Control

When a low data hiding bit-rate is demanded, distortion control is necessary in order to reduce the distortions introduced by the watermarking. A straightforward idea to control the distortions is to transform the pixel pairs only if they do not exceed a predefined error threshold. The error introduced by transforming the pixel $x$ is: $x' - x = 2x - y - x = x - y$. Similarly, the error introduced by transforming the other pixel of a pair is $y' - y = y - x$.

Let $\delta$ be the predefined error threshold. It immediately follows that this distortion control mechanism introduces a supplementary constraint: the pair $(x, y)$ is transformed if

$$|x - y| < \delta. \tag{5}$$

Equation (5) defines a strip-shaped domain, $\Delta$, located along the diagonal of $[0, L] \times [0, L]$ set. Therefore, RCM domain becomes the set intersection of $D$ and $\Delta$, $D \cap \Delta$ [see Fig. 1(b)].

By applying error control using the marking and detection algorithms of Sections III-A and III-B, it only means to update the transform domain (including the elimination of odd pairs subject to the ambiguity) for accounting for the error threshold. This corresponds to a slight mathematical complexity increase by verifying (5) for each pair to be transformed.

## IV. VERY LOW COMPLEXITY IMPLEMENTATION

In the marking stage, for each pair of pixels, the forward RCM needs two multiplications by two (in fact, simple arithmetical shifts) and two subtractions. This means one multiplication and one subtraction per pixel. No more than two comparisons per pixel are necessary to verify (2). Besides, there are some low-cost logical and bit manipulation operations. The detection stage is almost of the same complexity. The cost of the inverse RCM is slightly higher: for each pixel, one needs: one multiplication, one division, one addition, and one rounding to the integer value.

While the cost evaluated above appears as being very low, the complexity can still be drastically decreased. It is enough to observe that all the computations are performed on pairs of integers ranging in $[0, L]$. Instead of doing the actual calculation each time for each pair of integers, the idea is to look up for the answer in a memory table. The computations are performed once for all the possible pairs, and the results are stored in a lookup table (LUT) to be used for the marking/detection of all the images having the graylevel range in $[0, L]$.

For eight-bit pixels, the LUT for marking has $256 \times 256$ entries. Each entry contains the resulted pixels (16 bits) plus two extra bits, $F$ and $S$. $F = 1$ indicates that the LSB of the second pixel is available for data embedding, and $S = 1$ indicates that the LSB of the first pixel must be saved. The entry corresponding to the pair $(x, y)$ is located at the address $\overline{xy} = (L+1)x + y$. If $(x, y) \in D_c$ and $x, y$ are not both odd numbers, the entry contains the transformed pixels $x', y'$ with the LSB of $x'$ set to "1," $F = 1$, and $S = 0$. If $(x, y) \in D_c$ and $x, y$ are odd numbers, the entry contains the same values for $x, y$ (with the LSB of $x$ set to "0") and $F = 1$, $S = 0$. Finally, if $(x, y) \notin D_c$, the entry contains the original $x, y$ values (with the LSB of $x$ set to "0"), $F = 0$ and $S = 1$.

The LUT for detection is similarly organized. The meaning of $F$ and $S$ is slightly different. $F = 1$ indicates that a bit of data is embedded into the LSB of the second pixel ($F = 0$ if not). $S = 1$ indicates that the true LSB of the first pixel should be restored from the saved bit sequence, and $S = 0$ means that both pixels are already recovered. The entries corresponding to transformed pairs contain the recovered values computed by the inverse transform, $F = 1$ and $S = 0$. For the marked pairs of odd numbers, the LUT contains the original odd pixels values together with $F = 1$ and $S = 0$. For the pixel pairs that do not carry any embedded data, the entry contains the true seven bits of the first pixel, the entire second pixel, $F = 0$ and $S = 1$.

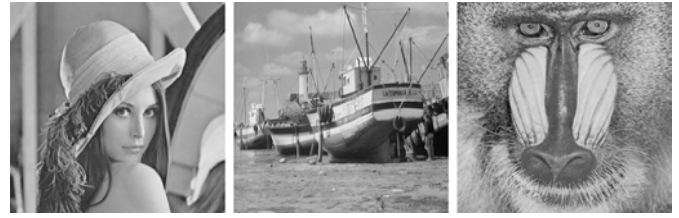

Fig. 2. Test images.

Obviously, the LUT implementation of the distortion controlled watermarking proceeds identically with the only difference being that the domain $D$ is replaced by $D \cap \Delta$. By using an LUT implementation, the marking reduces to simple memory addressing and some bits manipulation to collect the LSBs (according to the values of $S$) and to store them together with the watermark by overwriting the LSBs of the pixels pointed by $S$. The detection and original recovery is of the same complexity: watermark extraction from the pixels pointed by $F$, original values completely recovered from the table, or to be completed with the LSBs extracted from the watermark sequence (as pointed to by $S$). The size of the LUTs is affordable: for eight-bit pixels, it is slightly larger than a $512 \times 512$ graylevel image. For the general case of pixels ranging in $[0, L]$, one has two memory tables of $l^2$ entrys of $2l + 2$ bits, where $l = \lceil \log_2(L+1) \rceil$.

## V. EXPERIMENTAL RESULTS

The proposed scheme was tested on several graylevel and color images. Results obtained for the classical test images shown in Fig. 2 are presented. Applying the proposed scheme on *Lena* without control distortion, a bit-rate of 0.499 bpp is obtained. The bit-rate is very close to the theoretical upper bound, 0.5 bpp. Further iterations of the scheme increase the hiding bit-rate at 0.98, 1.40, 1.73, and 1.86 bpp. Details of four marked copies are shown in Fig. 3. For low and medium bit-rates, a slight increase of contrast can be seen. Increasing the hiding capacity, the noise increases as well.

The bit-rate obtained for the test image *Boat* is slightly lower: the maximum hiding capacity is of 1.53 bpp. This is explained by the fact that the *Boat* image has more details than *Lena*. The highly textured test image *Mandrill* provides only 0.84 bpp embedding rate. The PSNR of the test images with respect to the hiding bit-rate using control distortion is plotted in Fig. 4. The threshold was increased by a fine step of two graylevels. The distortion control scheme is implemented by the LUT as well.

In terms of data hiding bit-rate, the proposed reversible watermarking largely outperforms the compression-based schemes [1], [2]. The Celik's scheme [2], which appears to provide the highest bit-rates among the compression-based methods, provides a bit-rate of 0.17 bpp for *Mandrill* and of 0.68 bpp for *Lena* and *Boat* images.

The results are very close to Tian's [3]. The Tian's difference expansion scheme performs slightly better: the maximum bit-rate obtained for *Lena* is of 1.97 bpp, i.e., 5% higher. The difference expansion scheme is based on an integer wavelet transform defined on groups of two pixels. One bit of information is inserted into each transformed pixel pair, and then the inverse transform is performed. A location map is necessary to identify the pairs of pixels where information was inserted. The scheme provides at most a single bit of embedded data per pixel pair, i.e., a capacity of at most half of the size of the LSB plane. Since the size of the map is exactly half of the LSB plane, an additional lossless compression stage is mandatory. As it can be seen, both difference expansion and RCM schemes work on pairs of pixels. The RCM scheme works directly in the spatial domain, while Tian's scheme works in the transformed domain. Instead of a single transform stage, Tian's scheme should perform a pair of transforms, both for marking and detection. A similar increase of image contrast results for Tian's scheme, but the distortions introduced by the marking are lower. For instance, at 0.24, 0.46, and 1.44 bpp, Tian's scheme outperforms our scheme with 1.42, 2.87, and 3.42 dB, respectively. The results reported by Tian for the test image Lena are plotted as well in Fig. 4 (star symbols). The complexity is considerably higher for the Tian's scheme when compared with the RCM.

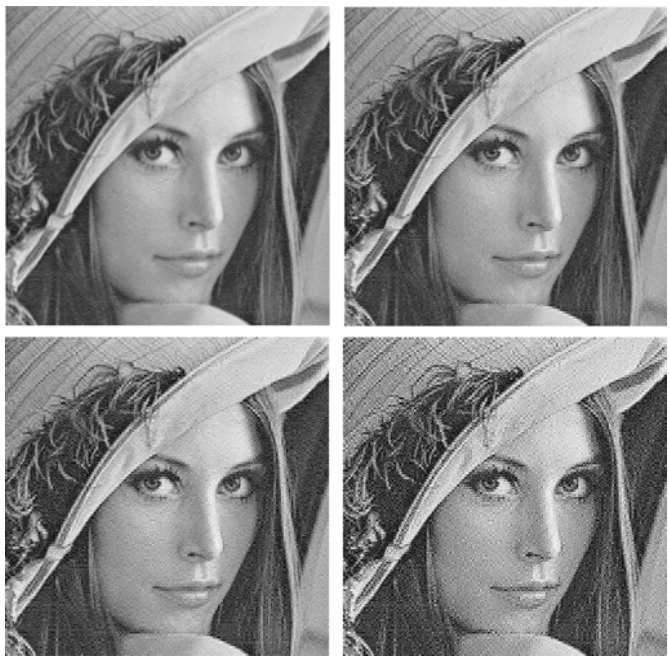

Fig. 3. Details of the marked copies in clockwise order: 0.40 bpp, 37.17 dB; 0.90 bpp, 28.48 dB; 1.30 bpp, 23.54 dB; 1.80 bpp, 16.23 dB.

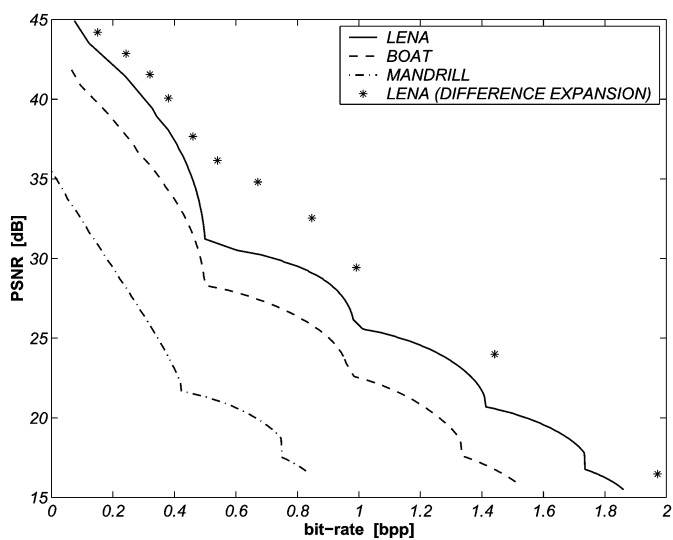

Fig. 4. PSNR with respect to the embedding bit-rate for the three test images.

An extended version of the difference expansion scheme transforms groups of more than two pixels [4]. Thus, the size of the location map decreases. The RCM scheme extends as well to groups larger than two pixels [7]. The extended RCM scheme can provide, in a single iteration, more than 0.5 bpp embedding rate (for instance, 0.68 bpp are obtained for the test image *Lena*). While the mathematical complexity of the extended RCM scheme remains very low, the LUT implementation becomes costly because of the size of the memory tables.

## VI. CONCLUSIONS

A spatial domain reversible watermarking providing high data embedding bit-rate at a very low mathematical complexity has been discussed. The proposed scheme does not need additional data compression. In terms of embedding bit-rates, the proposed scheme largely outperforms most of the reversible watermarking schemes reported in the literature and provides almost the same bit-rate as the difference expansion scheme and its extensions. In terms of mathematical complexity, the proposed reversible watermarking appears as being the lowest complexity scheme proposed so far. The computational complexity is reduced for both marking and decoding by using LUT access for each pair of pixels and some low complexity bit manipulation. This makes our scheme very appropriate for real-time applications. Finally, by distributing the location map and by storing the saved true values close to the corresponding pixel pairs, the RCM scheme provides robustness against cropping.